%% file: azzurri.tex
\documentclass[a4paper]{jpconf}
\usepackage{graphicx}
\usepackage{amsmath}
\usepackage{xspace}
\usepackage{pennames-pazo}
\input{ptdr-definitions}

\begin{document}

\title{First Results of Searches for New Physics at $\sqrt{s}$=7\;TeV with the
CMS detector} 

\author{Paolo Azzurri}

\address{Scuola Normale Superiore, Piazza dei Cavalieri 7, 56126 Pisa, Italy}

\ead{paolo.azzurri@cern.ch}

\begin{abstract}
First searches for new physics phenomena using the LHC 7 TeV proton-proton 
collision data collected by the CMS detector in 2010 are reviewed.
Results are presented of searches for new physics in events with hadronic jet pairs, 
and for heavy stable charged particles, including a dedicated search for 
long-lived particles that stop in the detector and 
decay in periods between beam crossings.
\end{abstract}

\section{Introduction}
During the 2010 LHC running, the CMS experiment~\cite{cms} recorded 
over 40~pb$^{-1}$ of pp collisions at 7~TeV center-of-mass energy.
The large momentum transfers accessible at these collision energies 
allow to test for new short range physics scenarios beyond the predictions 
of Quantum Chromodynamics (QCD) in dijet events, and to search for the 
production of new heavy quasi-stable particles~\cite{Fairbairn:2006gg}.

In the following, results for the search of new physics in dijet events 
are presented for resonances in the dijet mass spectrum
in Section~\ref{sec:resonances}, and for new interactions 
affecting dijet angular distributions 
in Sections~\ref{sec:centrality} and \ref{sec:angular}.
The outcome of searches for generic heavy stable charged particles
are given in Section~\ref{sec:hscp}, while the pursuit of 
long-lived gluinos that stop and decay in the CMS detector 
is detailed in Section~\ref{sec:gluinos}.

\section{\label{sec:resonances}Narrow dijet resonances}
In proton-proton collisions QCD predicts that the invariant mass spectrum of the two 
jets with largest \PT falls steeply and smoothly, while new massive objects
that couple to quarks ($q$) and gluons ($g$) would show up as resonant structures
in the dijet mass.
A data sample of  $2.9 \pm 0.3$~pb$^{-1}$ is employed for this analysis~\cite{resonance}, 
using a single-jet trigger with a transverse energy threshold of 50~GeV,
that is measured to be over 99.5\% efficient for dijet masses above 220~GeV.

\begin{figure}[htbp]
\begin{minipage}{0.49\linewidth}
\includegraphics[width=\columnwidth]{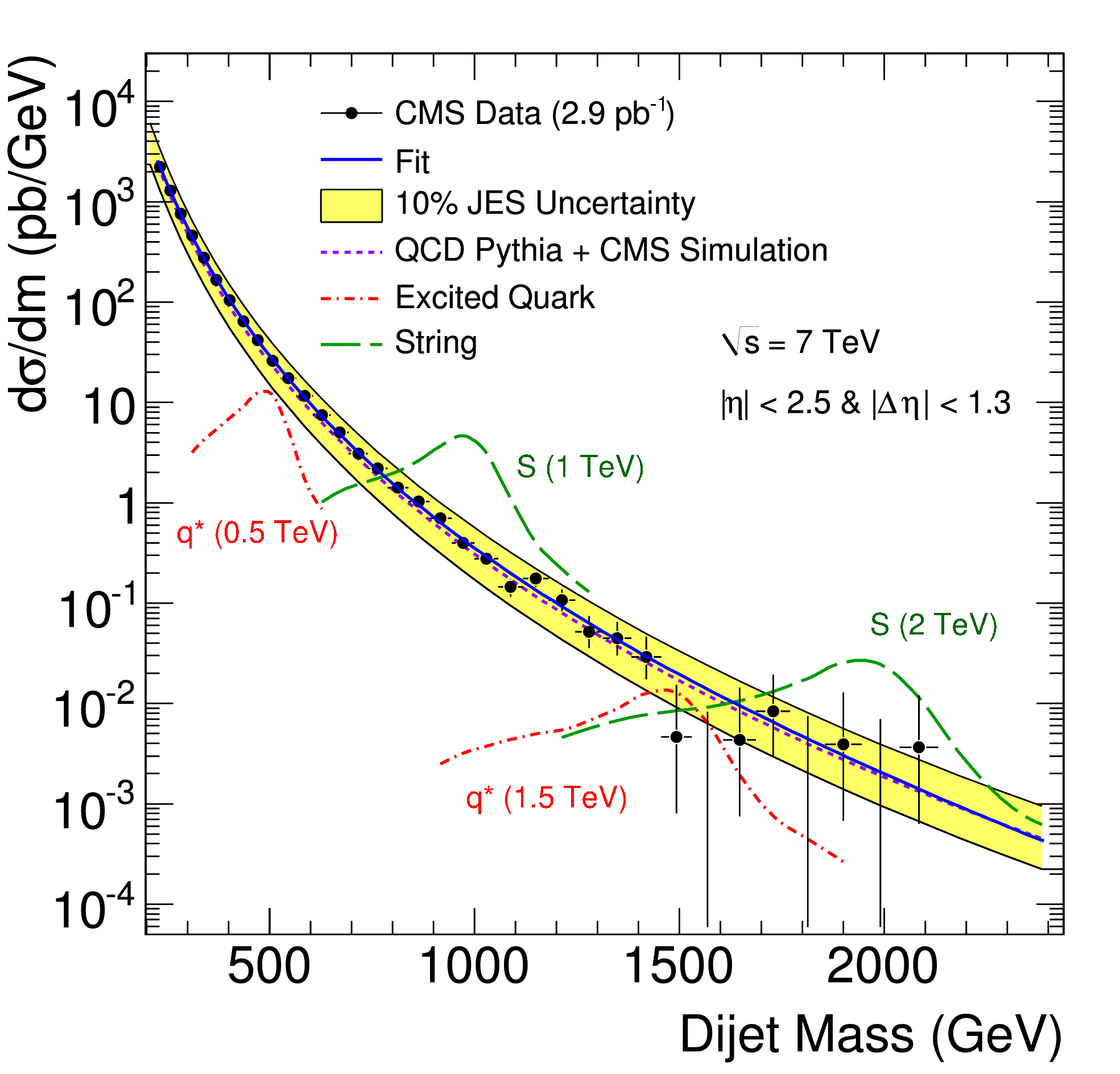}
\caption{\label{fig:mass} Dijet mass spectrum. The data points are superimposed 
with a smooth fit (solid line) and to QCD predictions~\cite{refPYTHIA} (short-dashed line).
The shaded band shows the jet energy scale uncertainty.
Possible excited quark signals and string resonance signals are also shown. }
\end{minipage}\hspace{1pc}%
\begin{minipage}{0.49\linewidth}
    \includegraphics[width=1.0\columnwidth]{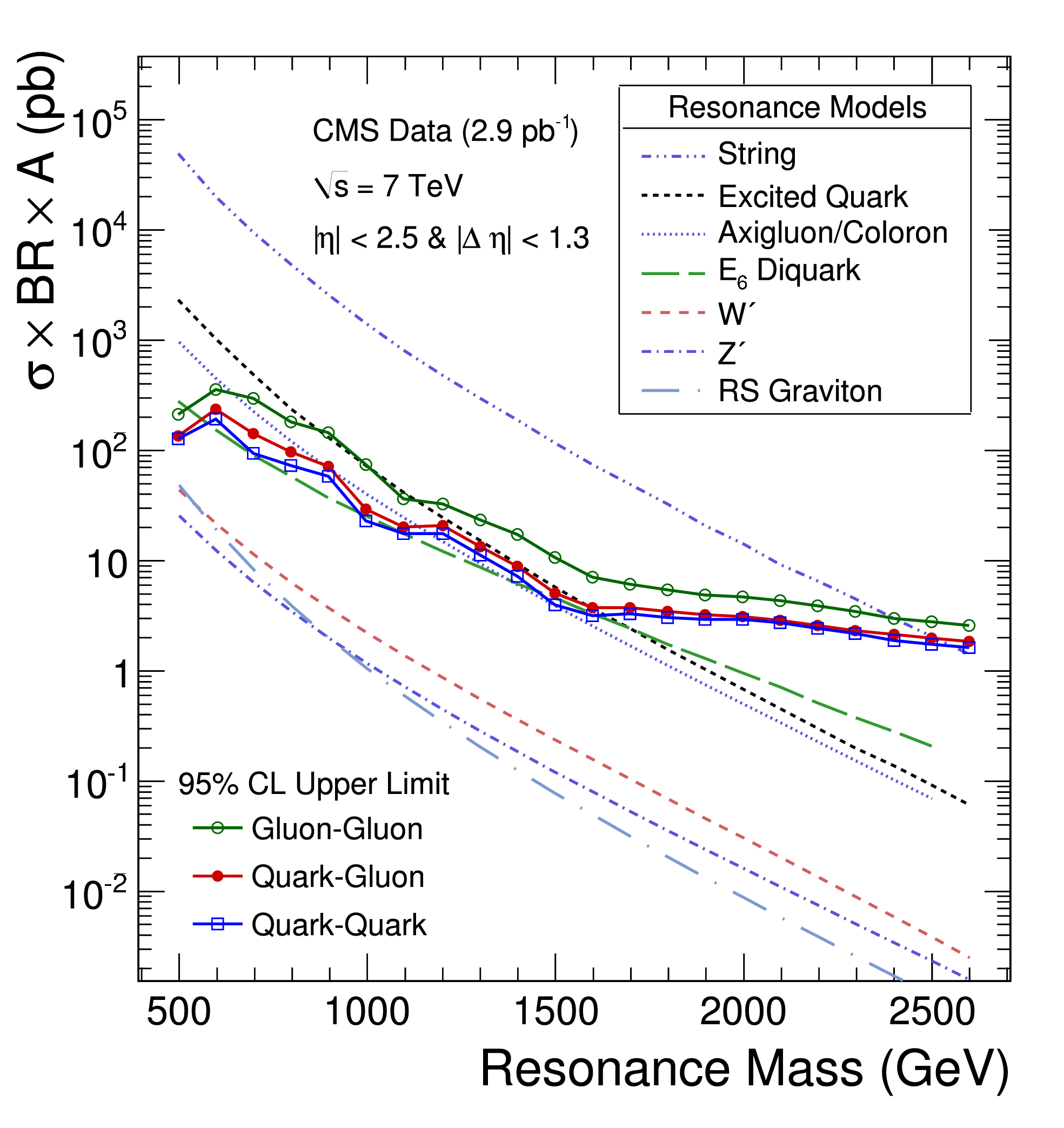}
    \caption{\label{fig:reslimits} 
Production upper limits for resonances  
decaying to gluon-gluon, quark-gluon, 
and quark-quark dijets, compared to theoretical predictions 
from different models.}
\end{minipage} 
\end{figure}

Jets are reconstructed with the anti-$k_T$ algorithm~\cite{1126-6708-2008-04-063} 
and $R=0.7$. The reconstructed jet energy $E$ is corrected as a 
function of \PT and $\eta$ for the
non-linearity and inhomogeneity of the calorimeter response~\cite{JME-10-003-PAS}.
The two jets with the leading \PT define the dijet system. The inclusive dijet mass 
is shown in Figure~\ref{fig:mass}, where the data points are compared with a prediction
from PYTHIA~\cite{refPYTHIA}, that is in agreement within the 10\% jet energy scale uncertainty.
The data is also compared with a smooth fit to a functional form with four free parameters,
and there is no indication of narrow resonances.

The dijet mass data points, the fitted background parameterization, and the expected 
dijet resonance shapes are used to set specific Bayesian limits on new particles decaying 
to parton pairs. 
The obtained 95\% confidence level (CL) upper limits on the resonances production 
are shown in Figure~\ref{fig:reslimits}
and compared to the expected productions in different models. 
The resulting excluded mass ranges for string 
resonances~\cite{Anchordoqui:2008di}, excited quarks~\cite{ref_qstar}, 
axigluons~\cite{ref_axi}, colorons~\cite{ref_coloron} and 
$\mbox{E}_6$ diquarks~\cite{ref_diquark} all extend previous 
exclusions~\cite{resonance}, while 
for new gauge bosons $W^{\prime}$ and 
$Z^{\prime}$~\cite{ref_gauge}, and Randall-Sundrum gravitons~\cite{ref_rsg}
no mass interval is excluded with the present data.

\section{\label{sec:centrality}Dijet centrality ratio}
In proton-proton collisions the QCD dijet production is dominated by t-channel processes
and peaks in the forward direction (along the beam axis) while new physics scenarios, 
including models of quark compositeness, predict more isotropic angular distribution, 
leading to enhanced jet production at smaller pseudorapidity values.

\begin{figure}[hbtpxs]
\begin{minipage}{0.49\linewidth}
\includegraphics[width=\columnwidth]{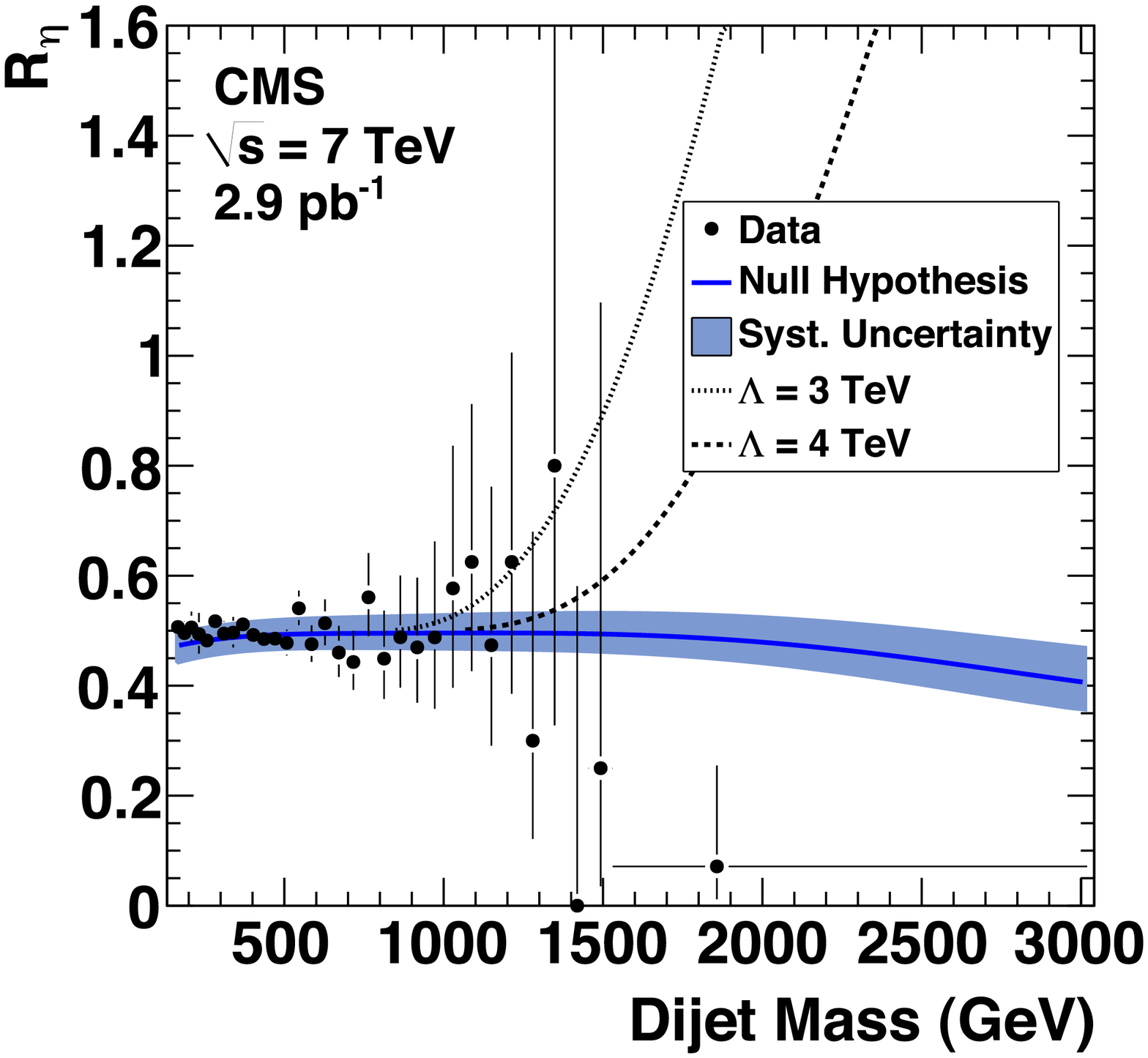}
\caption{The observed dijet centrality ratio as a
    function of $m_{jj}$ compared with QCD predictions
    (solid line), with the systematic uncertainty
    band, and to two hypotheses of quark contact interactions
    with $\Lambda = 3,4$ TeV.}
\label{fig:contact}
\end{minipage}\hspace{1pc}%
\begin{minipage}{0.49\linewidth}
\begin{center}
\includegraphics[width=\columnwidth]{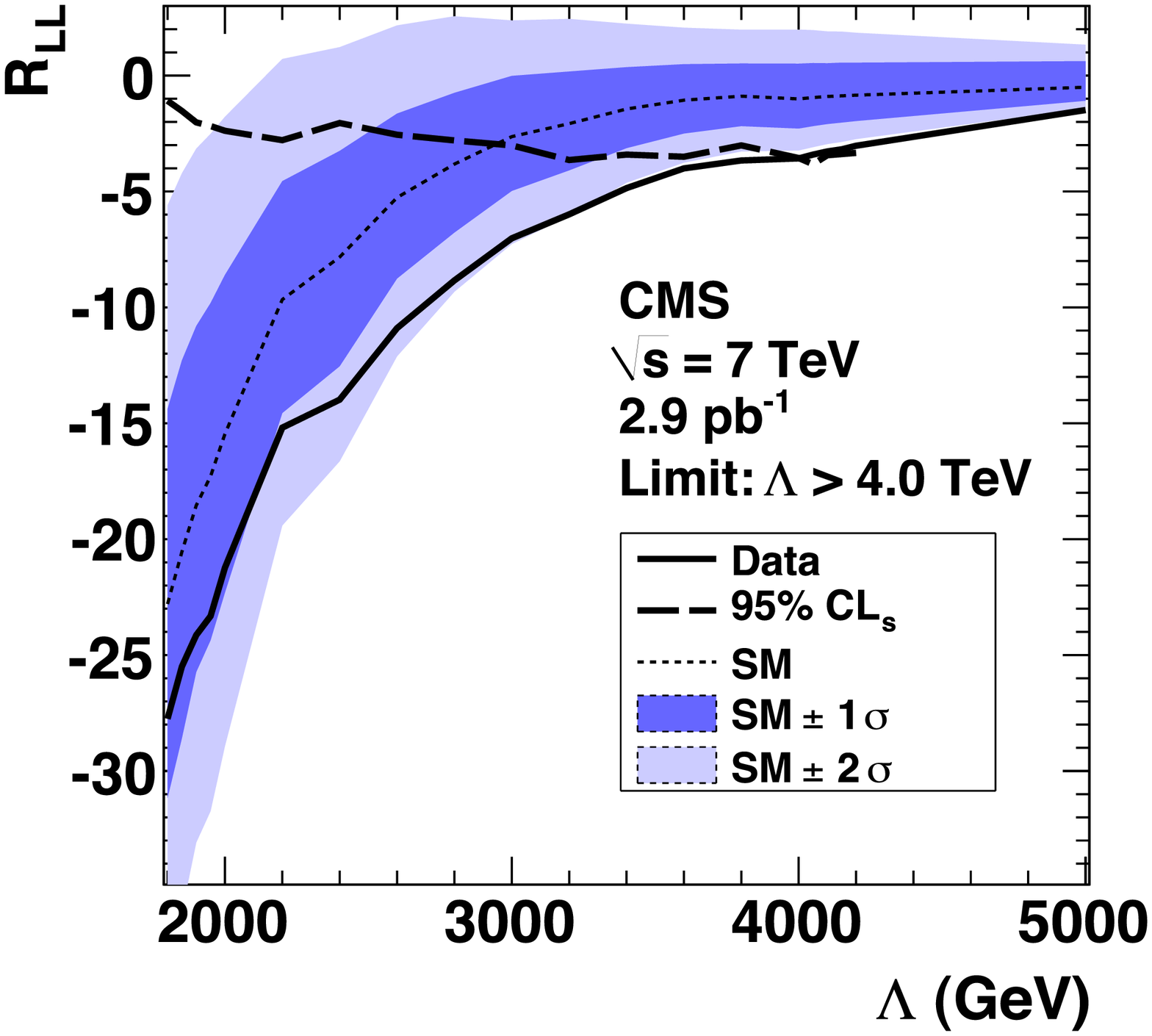}
\caption{Exclusions for the contact interaction
scale $\Lambda$.  We show ${\cal R}_{LL}$ versus $\Lambda$ for the data
  (solid line), the 95\% CL$_s$ (dashed line), and the SM expectation
  (dotted line) with 1$\sigma$ (dark) and 2$\sigma$ (light) bands.}
\label{fig:clim}
\end{center}
\end{minipage}
\end{figure}

The dijet centrality ratio $R_{\eta}$ between the number of events
with the two leading jets in the region $|\eta|<0.7$
and the number of events with the two jets in the region
$0.7<|\eta|<1.3$, is defined to use less angular information than
a full dijet angular distribution analysis, but allows for fine binning in the dijet mass
and cancels many sources of systematic uncertainty, providing an accurate test of QCD 
and probe to new physics. 
The analysis~\cite{centrality} is based on the same
data sample, events selection and jet reconstruction
used for the previously described dijet resonances search~\cite{resonance}. 
Figure~\ref{fig:contact} shows the predicted and observed values of $R_{\eta}$
a a function of the dijet mass $m_{jj}$. 
The measured $R_\eta$ is nearly flat with a value of about $0.5$, 
as predicted by QCD, while the presence of quark contact interactions 
would lead to an increase of  $R_{\eta}$ above a value of
$m_{jj}$ that depends on the compositeness energy scale $\Lambda$.

A log-likelihood-ratio ${\cal R}_{LL}$ is used to test for the presence 
of quark compositeness, and, given the consistency of the data with the QCD
hypothesis, a modified frequentist
method with the ${\rm CL}_{\rm s}$ approach~\cite{cls},
 is employed to derive the $95\%$ CL limit $\Lambda>4.0$ TeV
, as displayed in Figure~\ref{fig:clim}

\begin{figure}[h]
\includegraphics[width=0.55\linewidth]{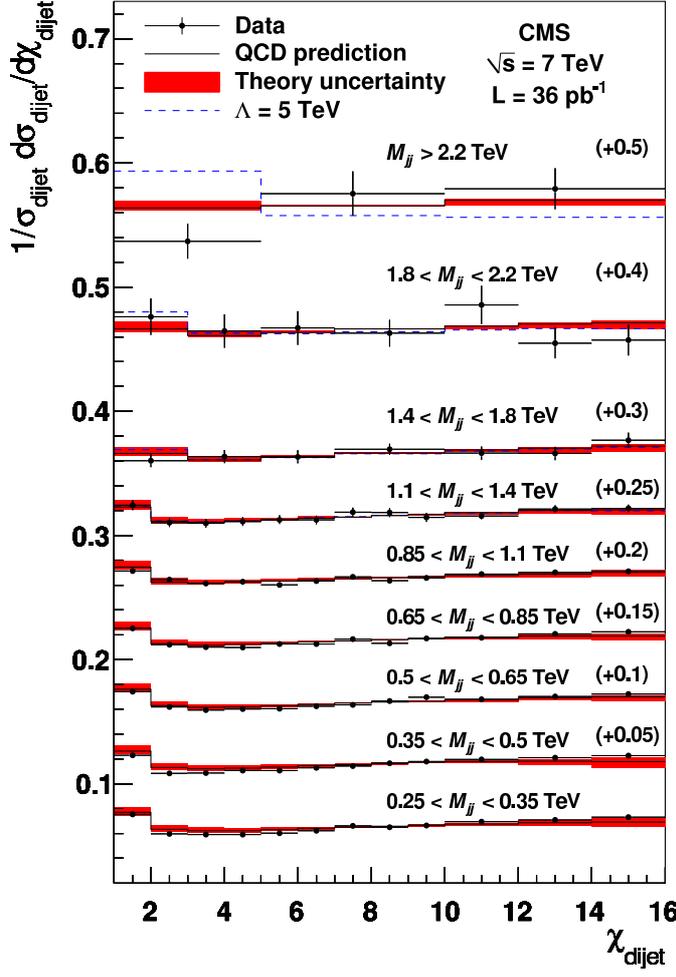}
\begin{minipage}[b]{0.35\linewidth}
\caption{\label{fig:chi}Normalized dijet angular distributions in
$m_{jj}$ ranges, shifted vertically (as given in parentheses) for clarity. 
The data points include full uncertainty bars and
are compared with the predictions of pQCD at NLO (solid
histogram) and with the predictions including a contact
interaction term of compositeness scale $\Lambda=5$~TeV (dashed
histogram). The shaded band shows the effect on the NLO pQCD
predictions due to factorization and renormalization 
scale variations and parton distribution functions 
uncertainties, as well as the uncertainties from the
non-perturbative corrections added in quadrature.}
\end{minipage}
\end{figure}

\section{\label{sec:angular}Dijet angular distributions}
The complete dijet angular distribution is measured 
though the differential cross section  
$(1/ \sigma_\text{dijet})(d\sigma_\text{dijet}/d\chi_\text{dijet})$
where $\chi_\text{dijet} = (1 +
|\cos{\theta^*}|) / ( 1 - |\cos{\theta^*}|)$,
and $\theta^{*}$ is the polar scattering angle in
the dijet center-of-mass (CM) frame~\cite{angular}.
This choice is motivated by the
fact that $d\sigma_{\rm dijet}/d\chi_{\rm dijet}$ is 
rather flat for QCD (Rutherford) parton scattering.

A data sample of up $36 \pm 4$~pb$^{-1}$ is employed for this analysis, 
and events are selected with five inclusive single-jet triggers, with
different jet transverse momentum \PT thresholds.
Offline jets are reconstructed with the anti-$k_T$ algorithm~\cite{1126-6708-2008-04-063} 
with $R=0.5$, and the reconstructed jet energy is corrected~\cite{JME-10-003-PAS}.
The differential dijet angular distributions for
different $m_{jj}$ ranges, normalized to their respective
integrals, are shown in Fig.~\ref{fig:chi}. The data are compared 
with perturbative QCD (pQCD) predictions at next-to-leading order (NLO),
for all $m_{jj}$ ranges.

\begin{figure}[ht]
\includegraphics[width=0.5\columnwidth]{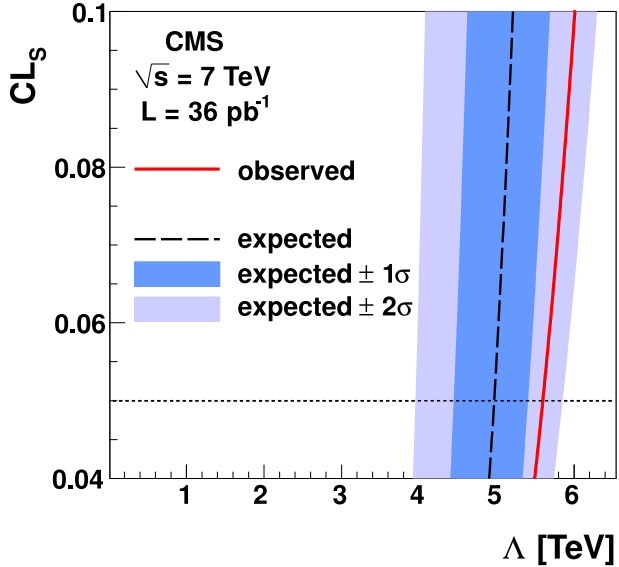}
\hspace{2pc}
\begin{minipage}[b]{0.4\linewidth}
\caption{\label{fig:chilim}The observed ${\rm CL}_{\rm s}$ (solid)
and expected ${\rm CL}_{\rm s}$ (dashed) with one (two)
standard deviation indicated by the dark (light) band as a
function of the contact interaction scale $\Lambda$, from the dijet 
angular distributions. The 95\%
confidence level limits on $\Lambda$ are extracted from the
intersections of the ${\rm CL}_{\rm s}$ lines with the horizontal 
line at ${\rm CL}_{\rm s}$=0.05.}
\end{minipage} 
\end{figure}

Given the good agreement with pQCD, the data are used to 
to set limits on quark compositeness represented by a four-fermion contact
interaction term in addition to the QCD Lagrangian.
As in the previous section, a statistical test 
based on the log-likelihood-ratio ${\cal R}_{LL}$ 
is used to derive limits on $\Lambda$, using the 
modified frequentist ${\rm CL}_{\rm s}$ approach~\cite{cls}.
The results are shown in Figure~\ref{fig:chilim}, and a 
95\% confidence level bound $\Lambda>$5.6~TeV is observed,
the currently most stringent limit on the contact interaction scale of
left-handed quarks.


\section{\label{sec:hscp} Heavy Stable Charged Particles}
Many extensions of the standard model predict the existence of 
heavy (quasi-)stable charged particles (HSCPs),
that would travel across the size of typical particle
detector~\cite{Fairbairn:2006gg},
and could be observable as high momentum ($p$) particles
with exceptional ionization energy loss ($dE/dx$).

\begin{figure*}[htbp!]
\begin{center}
\includegraphics[width=0.45\textwidth]{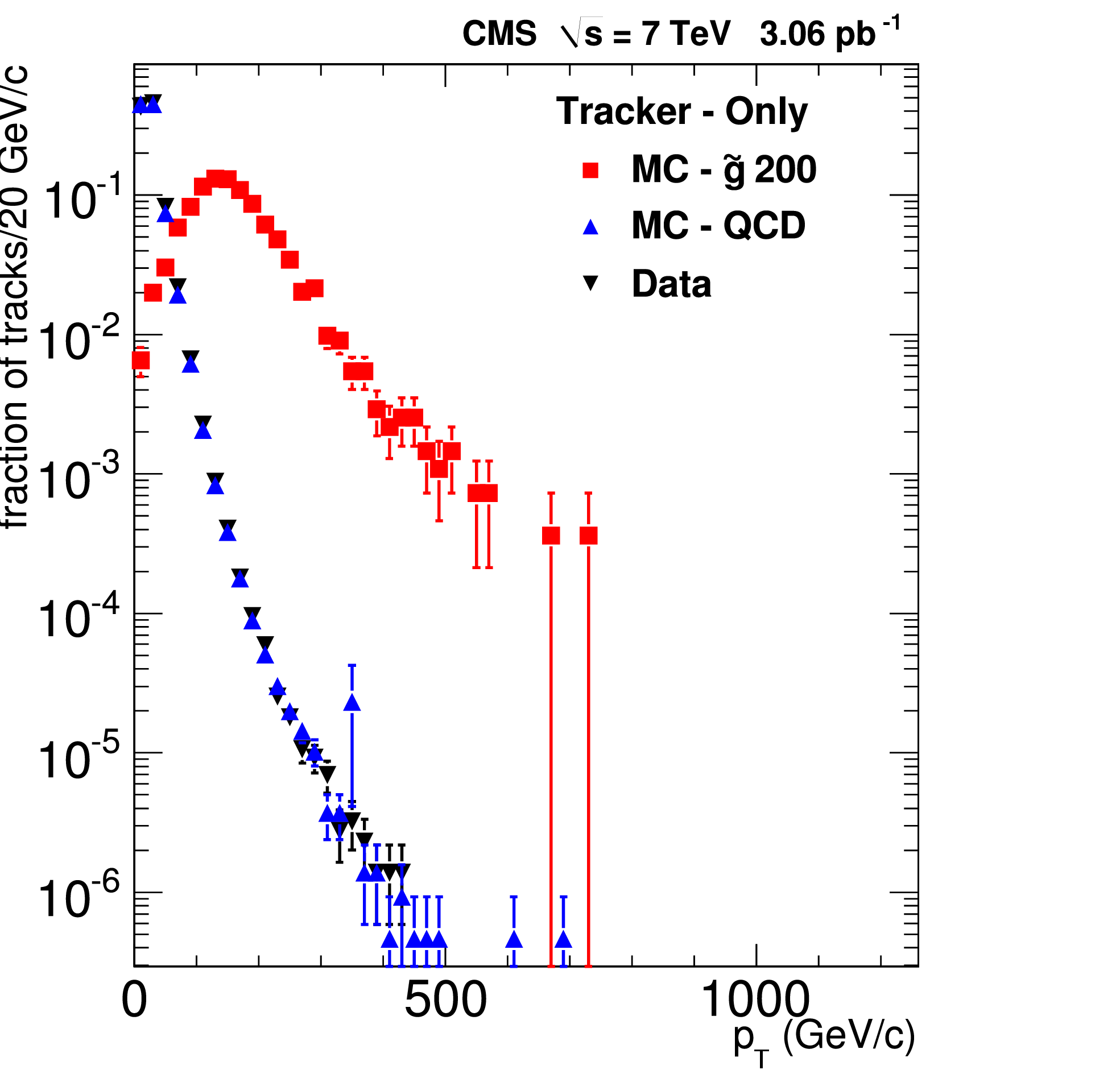}
\includegraphics[width=0.45\textwidth]{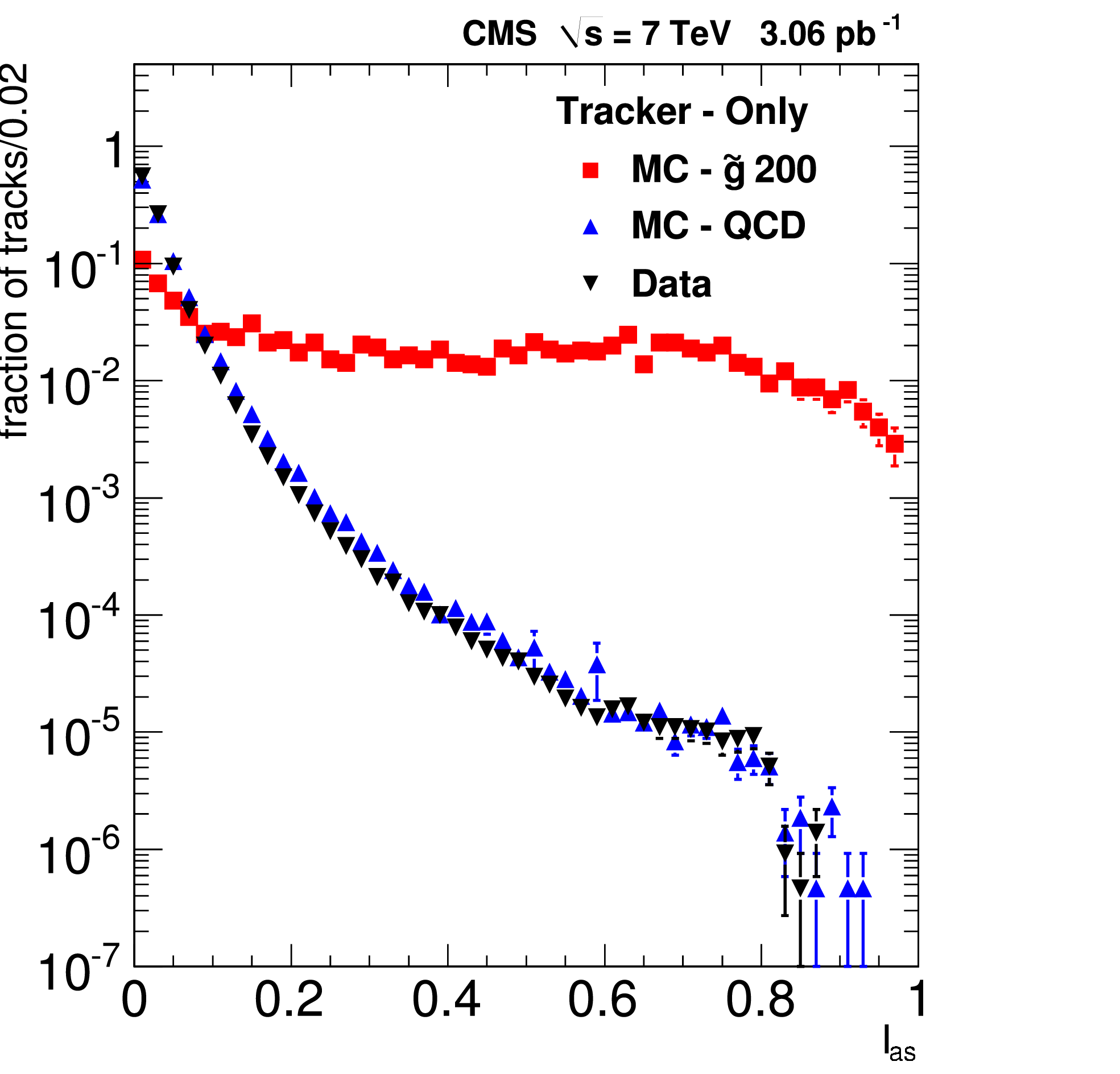}
\caption{Data normalized distributions of \PT (left) and $I_{as}$ (right),
compared to simulations from standard QCD processes 
and from pair-production of \PSg\  with a mass of 200 GeV/$c^2$.}  
\label{fig:hscp}
\end{center}
\end{figure*}

The CMS search~\cite{hscp} is based on an 
integrated luminosity of $3.1$ pb$^{-1}$,
and has two approaches. 
A ``\textit{tracker-only}'' selection looks for 
charged particles in the inner tracker 
 with large $dE/dx$ and \PT.
A ``\textit{tracker-plus-muon}'' selection 
additionally requires associated hits in the outer
muon detectors. 
For both selections, the mass of the candidate is
estimated from the measured $p$ and $dE/dx$.
For candidate tracks with $|\eta| < 2.5$ and
$p_T > 15$ GeV/$c$, an estimator $I_{as}$ of the degree of compatibility of the observed
charge measurements in the silicon detector 
with the minimum--ionizing particle (MIP) hypothesis is used 
to point out non-relativistic HSCP candidates.
The data and Monte Carlo (MC) normalized distributions of \PT and
$I_{as}$ for pre-selected tracker-only candidates are shown  in 
Figure~\ref{fig:hscp}. 
One MC sample contains events
from standard QCD processes, while the second MC sample contains signal events
from pair-production of stable \PSg\  with a mass of 200 GeV/$c^2$. Both
samples are generated with PYTHIA~\cite{refPYTHIA}.

The mass of the candidates is extracted from the measured $dE/dx$ and 
$p$ values, inverting a parametrized Bethe-Bloch fitted from data.
The search is performed as a counting experiment with candidates 
required to have $I_{as}$ and
$p_T$ greater than threshold values, and the
candidates mass to be in the range of 75 to 2000 GeV/$c^2$.

\begin{figure}
\vspace{-0.4cm}
\includegraphics[width=0.55\linewidth]{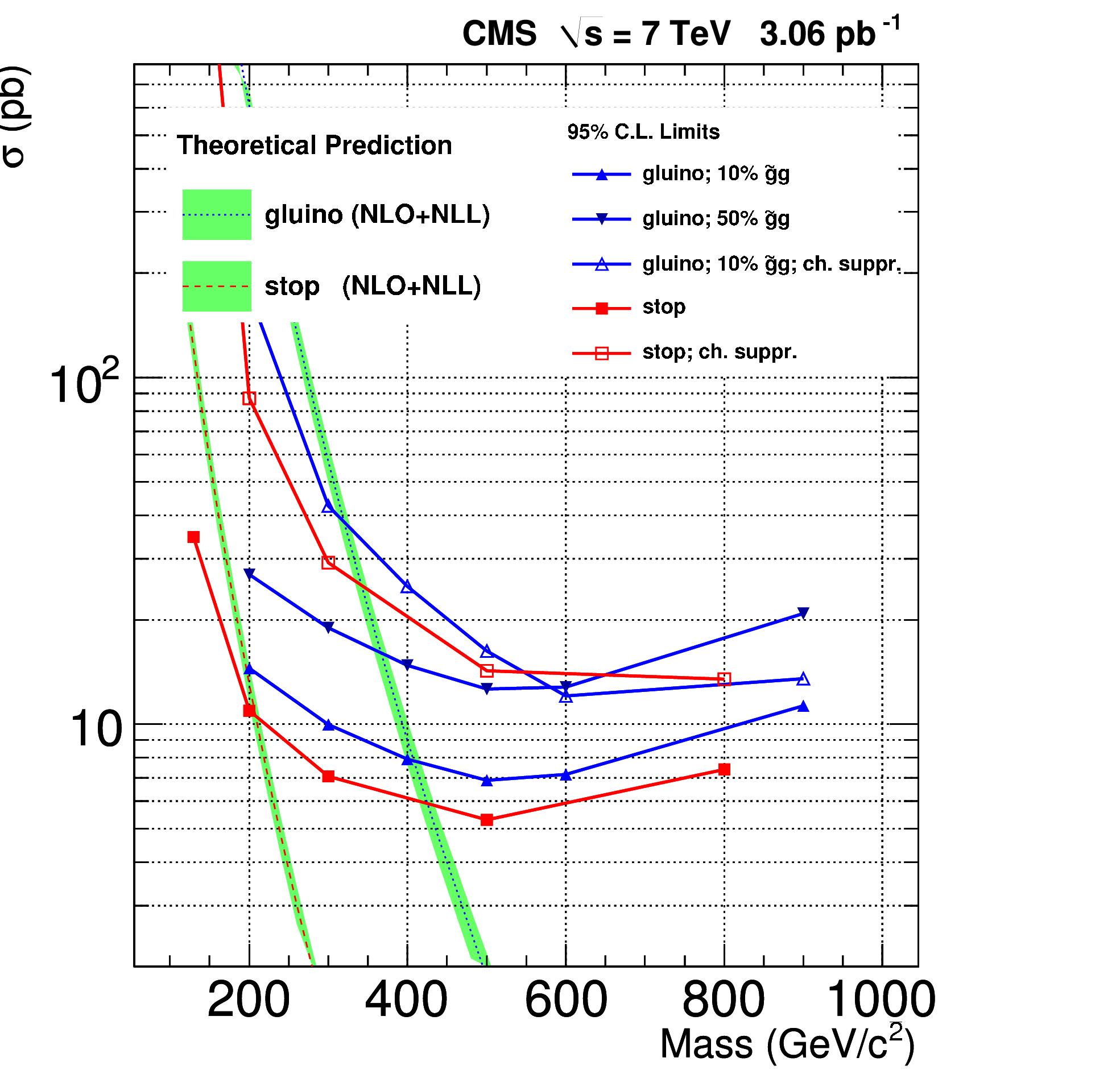}\hspace{2pc}%
\begin{minipage}[b]{0.4\linewidth} 
\caption{\label{fig:STOPMassExclusion} 
HSCP theoretical production cross section and observed 95\% CL upper limits 
for different combinations of models
and scenarios: pair production of supersymmetric stop and gluinos;
different fractions, $f$, of $R$-gluonball hadronization states  
and charge suppression scenarios~\cite{hscp}. Only the results
obtained with the most sensitive selection are reported: tracker-only
for the charge suppression scenarios and tracker-plus-muon
for all other cases.
The bands represent the theoretical 
uncertainties on the cross section values.  }
\end{minipage}
\end{figure}

No signal candidates are observed in the data, with $0.074 \pm 0.011$
expected form standard processes for the tracker-only selection, 
and $0.025 \pm 0.004 $ for the tracker-plus-muon selection.
Therefore cross section upper limits at the 95\%
CL are set on the HSCP production for two benchmark
scenarios: direct production of \PSg\ pairs and \stone pairs.
The signal efficiencies depend on the hadronization fractions 
and interactions of the resulting signal $R$-hadrons.
A summary of cross section upper limits, set with a
Bayesian method, are shown in Figure~\ref{fig:STOPMassExclusion}.
The tracker-plus-muon selection set a 95\% CL lower 
mass limit of 398 GeV/$c^2$ for long-lived \PSg\  with a $f=0.1$
fraction of $R$-gluonball states, and of 202 GeV/$c^2$ for 
long-lived \stone. 
In a pessimistic scenario of complete charge suppression 
the tracker-only selection yields a \PSg\  mass limit of 311 GeV/$c^2$.

\section{\label{sec:gluinos}Search for stopped gluinos}
A significant fraction of slow charged R-hadrons, 
resulting from the production of long-lived gluinos, 
can lose sufficient energy to be stopped inside the CMS detector
and may decay at later times (with $\tilde{g} \rightarrow g\tilde{\chi}^0_1$)
resulting in jet-like signals in the calorimeters, out-of-time with respect to LHC collisions.
The presence of a stopped-particle signal is searched for in 
 62 hours of trigger live-time, following collision data 
corresponding to 10~\pbinv recorded 
with $1 \times 10^{32}~{\rm cm}^{-2} {\rm s}^{-1}$ peak luminosity.
A control sample of 95 hours of trigger live-time
collected during fills with 
$2-7 \times 10^{27}~{\rm cm}^{-2} {\rm s}^{-1}$ luminosity is used 
to estimate the background rates~\cite{gluinos}.

\begin{figure}[hbt]
\hspace{-1.5cm}\includegraphics[width=0.85\linewidth]{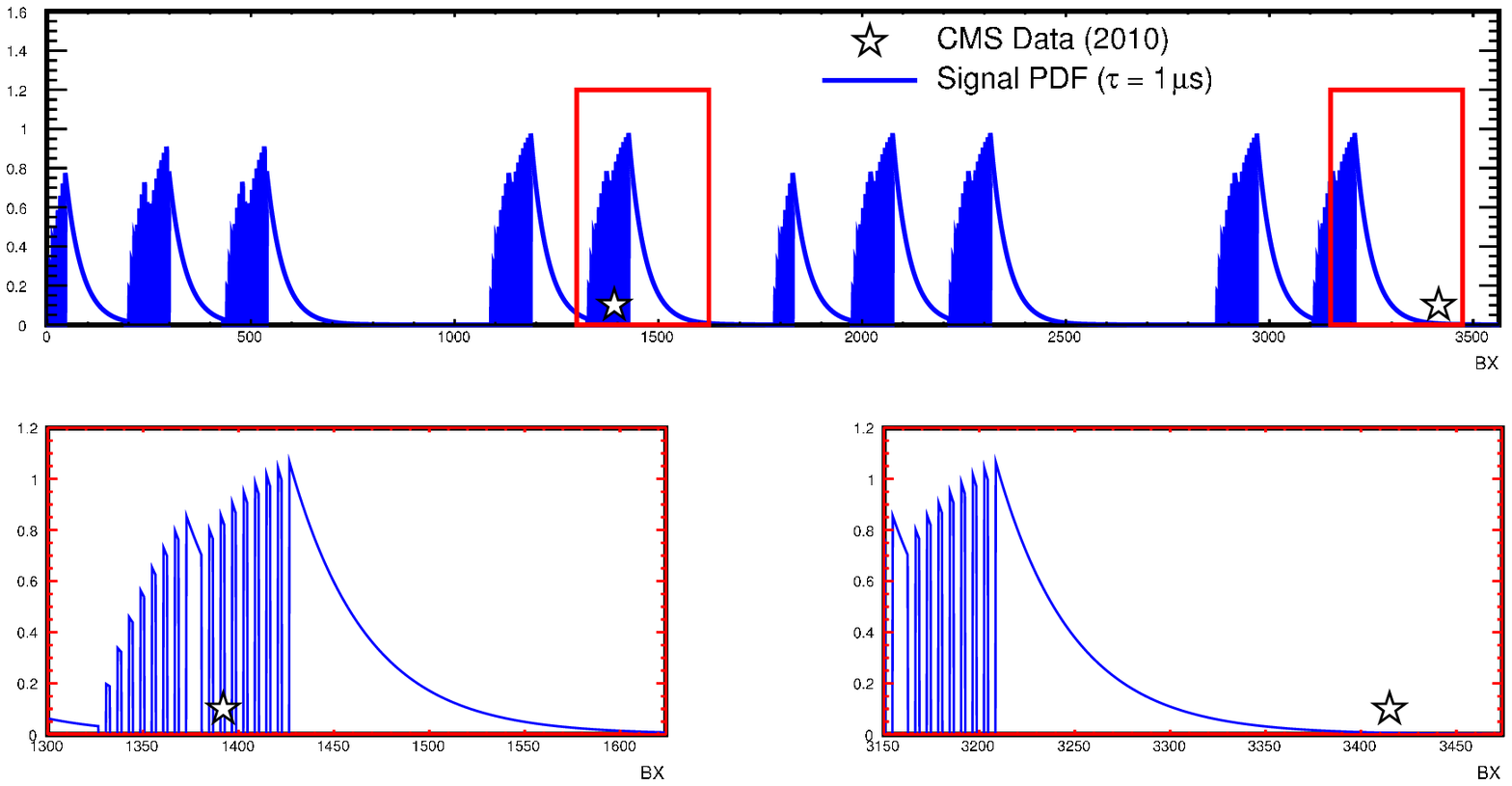}
\begin{minipage}[b]{0.22\linewidth} 
   \caption{In-orbit positions of two observed events recorded during an LHC fill with 140 colliding bunches. The decay profile for a 1 $\mu$s lifetime hypothesis is overlaid.  The bottom panels are zoomed views of the boxed regions around the two events. \label{fig:inorbitPDFsignal}}
\end{minipage}
 \end{figure}

A dedicated trigger is used that requires a jet-like signal
 when no beam is crossing CMS, and outside a 75~ns 
window around beam crossings.
Care is taken to reject beam-halo events, cosmic rays and calorimeter 
instrumental noise.
Both a counting experiment and a time-profile analysis are performed
for candidate events, building a probability  
density function (PDF) for the gluino decay signal. 
Figure~\ref{fig:inorbitPDFsignal} shows an example of such a PDF 
for a 1~$\mu$s lifetime, together with the in-orbit positions 
of two candidate events.

\begin{figure}[ht]
\begin{center}
\hspace{-1cm}\includegraphics[width=0.75\linewidth]{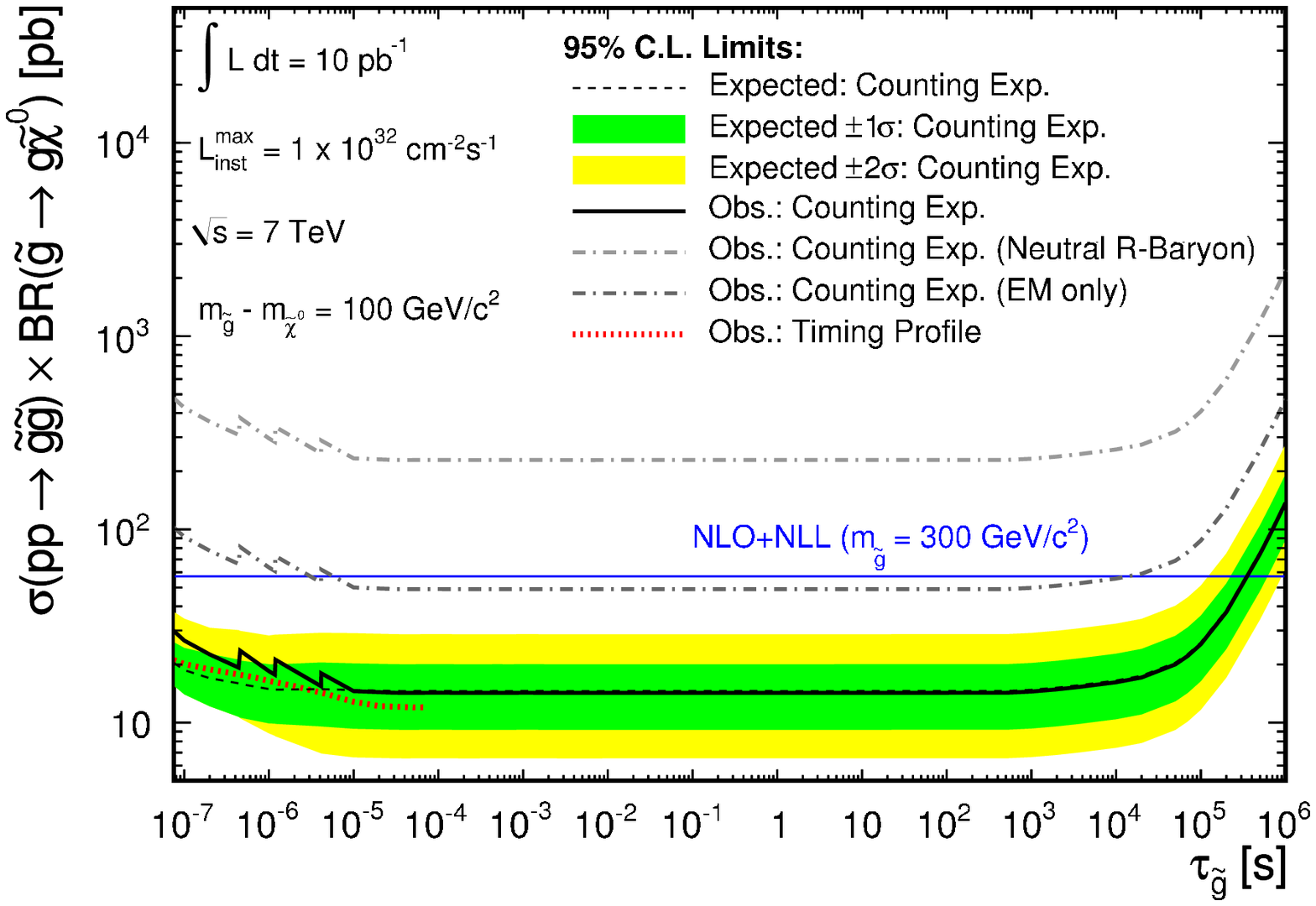}
\begin{minipage}[b]{0.28\linewidth} 
  \caption{Expected and observed 95\% CL limits on gluino pair production 
cross section times branching fraction as a function of the gluino 
lifetime from the counting experiment and the time-profile analysis. 
Alternative R-hadron interaction scenarios are also presented.  
The NLO+NLL production expectation for $m_{\tilde{g}} = 300$ GeV/$c^2$ is also indicated.  \label{fig:limit2syst}}
\end{minipage}
\end{center}
\end{figure}

Finally no excess of signal events over the expected background rates
is observed in the search sample, for any lifetime hypothesis, so that 
95\%~CL limits are set on 
$\sigma(pp \rightarrow \tilde{g}\tilde{g})  \times BR(\tilde{g} \rightarrow g\tilde{\chi}^0_1) $ 
for a given mass difference $m_{\tilde{g}}-m_{\tilde{\chi}^0_1} >100$ GeV/$c^2$,
and are summarized in Figure~\ref{fig:limit2syst}.
For $m_{\tilde{g}} = 300$ GeV/$c^2$  lifetimes from 75~ns to $3 \times 10^{5}$~s 
are excluded with the counting experiment.

\begin{figure}[ht]
\begin{center}
\hspace{-1cm}\includegraphics[width=0.75\linewidth]{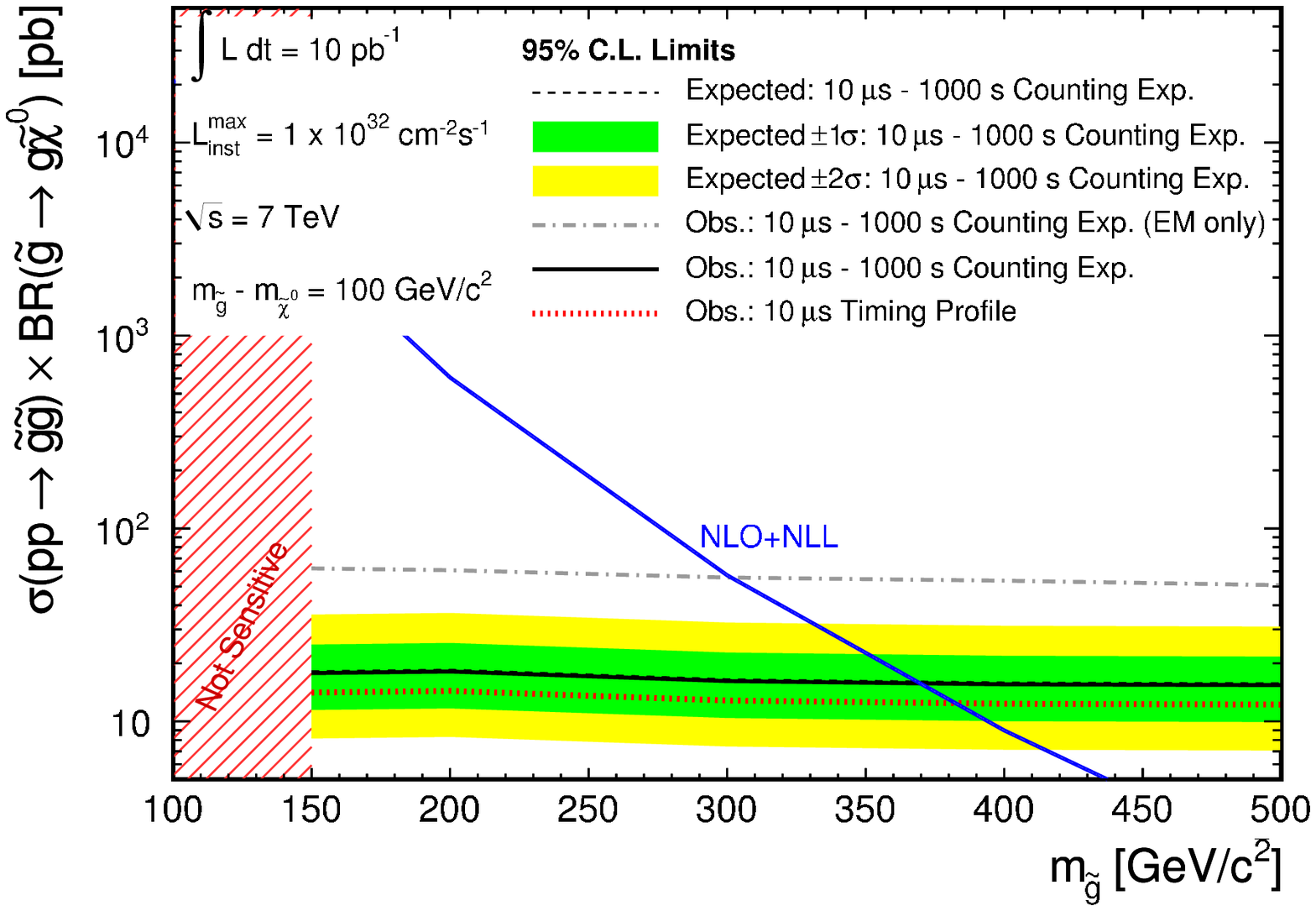}
\begin{minipage}[b]{0.28\linewidth} 
  \caption{Upper 95\% CL limits on gluino pair production 
cross section times branching fraction as a function of the gluino mass 
with strong or EM only   
R-hadron interactions. 
The $m_{\tilde{g}}-m_{\tilde{\chi}^0_1}$ 
difference is set at 100 GeV/$c^2$ and 
$m_{\tilde{\chi}^0_1} > 50$ GeV/$c^2$ is assumed.  
The NLO+NLL expected production are also shown.
The indicated lifetimes maximize the counting experiment and 
time-profile analysis sensitivity.
\label{fig:limitMass}}
\end{minipage}
\end{center}
\end{figure}

The excluded regions as a function of the gluino mass are plotted in Figure~\ref{fig:limitMass}.
For lifetimes between 10~$\mu$s and 1000~s $m_{\tilde{g}} < 370$ GeV/$c^2$ is excluded,
and if only electromagnetic (EM) interactions are considered the limit becomes 
$m_{\tilde{g}} < 302$ GeV/$c^2$.

\end{document}

%% file: ptdr-definitions.tex
\providecommand{\pbinv} {\mbox{\ensuremath{\,\text{pb}^\text{$-$1}}}\xspace}

\providecommand{\PT}{\ensuremath{p_{\mathrm{T}}}\xspace}

\providecommand{\stone}{\ensuremath{\tilde{\mathrm{t}}_1}\xspace}

\providecommand{\rpv}{\ensuremath{\rlap{\kern.2em/}R}\xspace}

